\documentclass[]{emulateapj}
\usepackage{multirow,natbib}

\shorttitle{In Search of the The Solar Family}
\shortauthors{Bland-Hawthorn, Krumholz, \& Freeman}

\newcommand{\ltsim}{\protect\raisebox{-0.5ex}{$\:\stackrel{\textstyle <}{\sim}\:$}}

\newcommand{\msun}{M_{\odot}}
\newcommand{\mstarmin}{M_{\rm *,min}}
\newcommand{\mstarmax}{M_{\rm *,max}}
\newcommand{\tsn}{t_{\rm SN}}
\newcommand{\tcr}{t_{\rm cr}}
\newcommand{\tform}{t_{\rm form}}
\newcommand{\tdiff}{t_{\rm diff}}

\newcommand{\msuncl}{M_{\odot,{\rm cl}}}

\newcommand{\kms}{\ km s$^{-1}$}

\begin{document}

\title{The long-term evolution of the Galactic disk traced by dissolving star clusters}

\author{ }
\affil{ }

\author{Joss Bland-Hawthorn\altaffilmark{1,2}}
\affil{Sydney Institute for Astronomy, School of Physics, University of Sydney, NSW 2006, Australia}
\altaffiltext{1}{Leverhulme Visiting Professor, Physics Department, University of Oxford, OX1 3RH, UK}
\altaffiltext{2}{Visiting Research Fellow, Merton College, University of Oxford, OX1 4JD, UK}
\email{jbh@physics.usyd.edu.au}

\author{Mark R. Krumholz}
\affil{Department of Astronomy \& Astrophysics, University of California, Santa Cruz, CA 95060, USA}

\author{Ken Freeman}
\affil{Research School of Astronomy \& Astrophysics, Mount Stromlo Observatory, ACT 2611, Australia}

\begin{abstract}
The Galactic disk retains a vast amount of information about how it came to be, and how it evolved over cosmic time. However, we know very little about the secular processes associated with disk evolution. One major uncertainty is the extent to which stars migrate radially through the disk, thereby washing out signatures of their past (e.g.\ birth sites). Recent theoretical work finds that such ``blurring" of the disk can be important if spiral arms are transient phenomena. Here we describe an experiment to determine the importance of diffusion from the Solar circle with cosmic time. Consider a star cluster that has been placed into a differentially rotating, stellar fluid. We show that all clusters up to $\sim 10^4$ $M_\odot$ in mass, and a significant fraction of those up to $\sim 10^5$ $M_\odot$, are expected to be chemically homogeneous, and that clusters of this size can be assigned a unique ``chemical tag" by measuring the abundances of $\lesssim 10$ independent element groups, with better age and orbit determinations allowing fewer abundance measurements. The star cluster therefore acts like a ``tracer dye", and the present-day distribution of its stars provides a strong constraint on the rate of radial diffusion or migration in the Galactic disk. A star cluster of particular interest for this application is the ``Solar family" $-$ the stars that were born with the Sun. If we were able to identify a significant fraction of these, we could determine whether the Sun was born at its present radius or much further in. We show all-sky projections for the Solar Family under different dynamical scenarios and identify the most advantageous fields on the sky for the experiment. Sellwood \& Binney have argued for strong radial transport driven by transient spiral perturbations: in principle, we could measure the strength of this migration directly. In searching for the Solar family, we would also identify many thousands of other chemically homogeneous groups, providing a wealth of additional information. We discuss this prospect in the context of the upcoming HERMES high-resolution million-star survey.
\end{abstract}

\keywords{Galaxy -- stellar populations -- open clusters -- elemental abundances}

\section{Introduction}

An axiom of chemical evolution models is that stars form {\it in situ} within the disk, or at least they have not 
migrated far from their formation radius. But it has long been recognized that the metal abundance scatter at all radii is very 
large and comparable to the amplitude of the overall declining slope in [Fe/H]. This led to the idea that either the ISM was much less 
homogeneous in the past or that stars have migrated over large radial distances (Wielen, Fuchs \& Dettbarn 1996).
The allowed distribution of orbit ellipticities accounts for only half of the scatter (e.g. N\"ordstrom et al 2004).
Radial migration requires a strong non-axisymmetric impulse, i.e. perturbations from massive clouds or spiral arms.
Cloud deflections are closely related to spiral arm impulses because the orbiting cloud generates a spiral response
(``swing amplifier'')
within the disk (Julian \& Toomre 1966). In a seminal paper, Sellwood \& Binney (2002; hereafter SB02) went on to show that transient
spiral perturbations can lead to strong radial ``churning'' in disk galaxies (i.e. mixing without heating), and this can 
reasonably explain the observed abundance scatter (Sch\"onrich \& Binney 2009).

In a frame that rotates with the pattern speed of the spiral arm, stars close to the co-rotation 
radius orbit a point moving at the pattern speed. Inside co-rotation, stars overtake the pattern and fall into the spiral potential; this boosts their angular momentum (measured about the rotation axis of the disk) which moves them to a larger radius.  The ejected star now lags behind the pattern speed and falls backwards into the wave approaching from behind. This lowers its
angular momentum and the star returns from whence it came which results in {\it no overall migration} for an individual star.
The key insight is to consider what happens if spiral arms are transient
(e.g. Barbanis \& Woltjer 1967). This results in a strong stellar migration with individual stars moving inwards and other stars moving outwards  -- the Sellwood-Binney mechanism. Stars near resonance can lose or gain a substantial fraction of their angular momentum in a single encounter.
This process has now been seen in N-body simulations (gas$+$stars) of evolving disks that are kept sufficiently 
cool from halo gas accretion (Ro\v{s}kar et al 2008; S{\'a}nchez-Bl{\'a}zquez et al 2009), and the radial churning appears to be strong.

In support of radial migration, the Sun appears to be +0.17 dex too metal rich for its environs (Edvardsson et al 1993) which led Wielen et al (1996) to suggest that the Sun has moved outwards by about 2 kpc over the course of its lifetime. SB02 note that the Sun's highly circularised orbit is consistent with having migrated to its present position through churning. All stellar populations, from the oldest (e.g. planetary nebulae; Maciel et al 2006) to the youngest (e.g. Cepheids; Andrievsky et al 2004), exhibit a marked abundance gradient and show evidence of having flattened with cosmic time. This provides some justification for associating a metal-rich Sun (cf. Asplund et al 2005; Haywood 2008) with a smaller birth radius. Interestingly, Ro\v{s}kar et al (2008) find that metal-rich stars, like the Sun, could have originated almost anywhere. They find that half of the stars within the Solar torus today have come from large radial distances ($\gtrsim$2 kpc).

So where are the stars that were born with the Sun? If we were able to identify a useful fraction of the 
Solar family,\footnote{The Solar family was first discussed by Bland-Hawthorn \& Freeman (2004): these siblings should 
not be confused with ``Solar twins'' (e.g. Hardorp 1978; Dorren \& Guinan 1994; Cayrel de Strobel 1996), i.e. stars that are essentially identical in character to the Sun but at different stages of their evolution. By definition, these cannot be members of the Solar family.} we would be able to answer this question. We can think of the Solar family as a Òtracer dyeÓ that has been dropped into a differentially rotating, stellar fluid.  If the Sun was born at its present radius, in the absence of strong radial scattering, we would assume 
that the Solar family is confined to the Solar torus. In this picture, the width of the annulus is likely to reflect the epicyclic radial amplitude ($\sim 1$ kpc) of stars with the age of the Sun (\S5.3). If the stars are found to lie along a thinner annulus, this would indicate possibly that the orbits have been confined by strong resonances from the central bar, say.
If most of the Solar family was found to lie inside of the Solar circle, we would infer that it has experienced a steady outward radial migration. Grenon (1999) considered the possibility that stars are formed near the Galactic centre and then diffuse outwards. If we were unable to find any siblings, this may present the biggest surprise of all (see \S 2).

The important issues of birth place and radial migration are difficult to unravel. Just how secular migration evolves depends in large part on
the Galaxy's accretion history which sustains and triggers the onset of spiral disturbances (Sellwood \& Carlberg 1984). The stars that are 
most affected by angular momentum transfer lie close to the corotation radius, but the stochastic nature of the spiral disturbances give rise to a range
of pattern speeds over the same time interval (Sellwood 2008). In particular, in the SB02 simulation, the power spectrum of the $m=2$ modes reveals a wide range of pattern speeds (their Fig. 10), with corotation radii in the range 4 kpc to 20 kpc.

If the Sellwood-Binney mechanism\footnote{These models treat only stellar migration. It is interesting to consider whether gas experiences the same churning or whether the short lifetime of 
giant molecular clouds ($\sim 30$ Myr; \citealt{fukui09a}) suppresses it.}
is as efficient as suggested by Ro\v{s}kar's models, we are forced to conclude that the conventional approach to chemical 
evolution modelling (cf. Chiappini et al 2001) must be overhauled (Sch\"{o}nrich \& Binney 2009). Furthermore,
we must embrace a broader range of chemical
elements in surveys involving millions of stars (see Bland-Hawthorn \& Freeman 2004). Most stars are born in a single burst within compact clusters and stellar fragments. We can use Òchemical taggingÓ to ascertain whether these substructures are homogeneous systems or not. De Silva et al (2006, 2007a,b) have recently demonstrated that open clusters are chemically homogeneous, and even moving groups can show the same signature (see also Castro et al 1999; Sestito et al 2007; Randich et al 2006; Shen et al 2005; Chou et al 2010). Apart from a few light elements, globular clusters are also highly homogeneous (Gratton, Sneden \& Carretta 2004). The fact that old clusters are chemically homogeneous argues strongly against internal stellar processes or external ISM accretion as influencing the chemical signatures. We can therefore reconstruct many stellar clusters through their unique chemical signatures.

It is now possible to obtain echelle spectroscopy ($R\gtrsim 30,000$) for millions of stars in campaigns of a few years. 
The technique of ``chemical tagging'' (Freeman \& Bland-Hawthorn 2002) is being addressed by ongoing and planned stellar surveys. 
These include the HERMES survey on the Anglo-Australian Telescope 
(Freeman \& Bland-Hawthorn 2008; Freeman, Bland-Hawthorn \& Barden 2010), and the APOGEE survey at Apache Point Observatory
(Allende Prieto et al 2008).

The structure of our paper is as follows. In \S2, we estimate the birthrate of clusters within the Solar torus as a function of mass.
In \S3, we examine their likely chemical uniformity as a function of the progenitor cloud mass. In \S4, we derive more stringent limits on the size of the Solar family at the time of its birth. In \S5, we simulate
four possible distributions for the Solar family and determine their projection on the sky.
In \S6, we describe an experiment aimed at finding members of the Solar family before a discussion of the implications
in \S7.

\section{Formation of Star Clusters}

If we consider our volume of interest to be the Solar torus, defined as the annulus 2 kpc thick in radius centered on the Solar circle at $R_o\approx 8$ kpc (see \S~\ref{Solarparam}), then the cluster birth rate within this region is
\begin{equation}
\label{eq:birth}
\frac{d^2 \mathcal{N}}{dt\, d\ln (M_*/\msun)}=  3 \times 10^3 \dot{M}_0 \left(\frac{M_*}{10^4\,\msun}\right)^{-1}\mbox{ Gyr}^{-1},
\end{equation}
where $\dot{M}_0 = \dot{M}/(\msun\mbox{ yr}^{-1})$ and $\dot{M}$ is the total star formation rate in the Galaxy and the upper and lower cluster mass limits are ($\mstarmin, \mstarmax$) $=$ ($50\msun, 2\times 10^5\msun$). Thus, for example, in the last Gyr we expect $\sim 10^4$ clusters whose masses are within a factor of three of $10^4$ $\msun$ to have formed within the Solar torus if the star formation rate over this time has been close to its present value of $\sim 3$ $\msun$ yr$^{-1}$ \citep{mckee97}. This result is quite insensitive to the assumed limits on the cluster mass function or the extent of the star-forming disk. Changing $\mstarmin$ or $\mstarmax$ by a factor of 10 alters the estimated cluster birth rate by only $\sim 20\%$.

All but a small fraction of clusters formed are unbound, and the stars born within them drift off to become part of the field star population. By comparing catalogs of open and embedded clusters, \citet{lada03} estimate that less than 10\% of clusters survive more than 10 Myr, and that only $4-7\%$ survive for 100 Myr\footnote{We point out, however, that the distribution must have a long tail because a few open clusters have ages $\gg  1$ Gyr (Friel 1995).}. Moreover, analysis of extragalactic cluster populations (albeit with masses significantly larger than the typical Milky Way cluster) suggests that the survival fraction is nearly independent of mass until two-body relaxation becomes the dominant destruction process at times $\gg 100$ Myr after cluster formation 
\citep{fall05a, fall09a, fall06a, whitmore07a}. During this first 100 Myr, roughly $80-90\%$ of clusters are disrupted independent of mass, while two-body relaxation at longer times preferentially destroys low mass clusters. Since we do not strive for better than 10\% accuracy in our prediction of the disrupted cluster birth rate, we will not introduce the added complication of a mass- or time-dependent disruption rate. Instead we simply assume that a fixed 95\% of all clusters that are born are disrupted, independent of mass and time. This will slightly overestimate the number of disrupted clusters that formed less than 100 Myr, but for a relatively quiescent star formation history such as the Galaxy's, these clusters constitute a negligible fraction of the total population. Thus, if we target stars older than 100 Myr, the birth rate of now-disrupted clusters is given by
equation \ref{eq:birth} to within 10\%.

\section{Cluster Chemistry}
\label{clusterchemistry}

\subsection{Introduction}

Since most stars are not born in clusters that remain bound, and thus do not end up in open clusters more than 10 Myr later, the clusters observed by De Silva and collaborators presumably represent a biased sample. They are either the innermost remnants of clusters that lost most of their members and produced only a small bound core \citep[e.g.][]{kroupa01b}, or they are clusters that remained bound because they had star formation efficiencies significantly higher than the typical $10-30\%$ \citep{lada03}. We would therefore like to define some rough theoretical expectations about the chemical homogeneity of the disrupted cluster population.

Mixing of heavy elements in protoclusters is governed by several time scales. The dominant mode of chemical mixing in molecular clouds is turbulent diffusion \citep{murray90a}. The time scale associated with this process is $\tdiff\sim H^2/K$ \citep{xie95a}, where $H=(H_i^{-1} - H_{\rm H_2}^{-1})^{-1}$ is a composite between the scale heights of the H$_2$ molecules, $H_{\rm H_2}=n_{\rm H_2}/|\nabla n_{\rm H_2}|$, and of the species $i$ whose diffusion time we are computing, $H_i=n_i/|\nabla n_i|$. Here $n_{\rm H_2}$ and $n_i$ are the number densities of the two species. The quantity $K$ is the diffusion coefficient, which is of order $\sigma L$, where $\sigma$ is the turbulent velocity dispersion and $L$ is the correlation length of the turbulence. Observations indicate that the turbulent motions in molecular clouds are correlated on the scale of the entire cloud \citep{heyer04}, so $L$ is roughly the size of a cloud. If we consider a species that is distributed with an inhomogeneity comparable to or greater than that of the molecular hydrogen, $H_i \ltsim H_{\rm H_2}$, and a composition gradient on the scale of the entire cloud, $H_i \sim L$, then $\tdiff \sim L/\sigma = \tcr$, where $\tcr$ is the cloud crossing time. Thus, the time required for turbulence to smooth out a large-scale chemical gradient in a protocluster gas cloud is of order the cloud crossing time; smaller scale gradients disappear more quickly, with the time required to smooth out the gradient varying as the square of its characteristic size scale.

Star formation in a molecular cloud also appears to begin in no more than a crossing time after its formation \citep[e.g.][]{tamburro08a}, so there is no clear separation in time scales between chemical homogenization and the onset of star formation. This suggests that molecular clouds cannot be too far from homogeneous when they are assembled, but that the homogenization need not be total, since small residual chemical gradients will be wiped out as star formation proceeds. Thereafter, changes in element abundances, and inhomogeneities in the resulting stars, may be produced by supernovae within the cloud. In principle, even a single supernova is sufficient to change the chemical signature measurably. For example, \citet{De-Silva07a} find that scatter in Fe abundance in the HR 1614 moving group is roughly $0.01$ dex. The Fe content of the Sun is $\approx 10^{-3}$ $\msun$ \citep{anders89a}, so the measured \citeauthor{De-Silva07a} scatter corresponds to a scatter in mass fraction of $\sim 10^{-5}$. A supernova from a 15 $\msun$ star produces $\sim 10^{-1}$ $\msun$ of Fe \citep{woosley95a}, so if that supernova occurs inside a star-forming cloud that continues forming stars thereafter, the change in iron abundance will be measurable even if the supernova ejecta are mixed with $10^4$ $\msun$ of pre-supernova gas. The time required for a very massive star to evolve from formation to explosion defines the supernova time scale, $\tsn \approx 3$ Myr. Thus, while the ``initial conditions" for cluster formation are well-mixed, we only expect star clusters to be homogeneous if they are assembled on time scales shorter than $\tsn$.

\subsection{Chemical homogeneity}

To determine under what conditions this requirement is satisfied, we must compare $\tsn$ to the cluster formation timescale $\tform$. There is considerable debate about whether star cluster formation occurs in a single crossing time \citep{elmegreen00, hartmann01, elmegreen07, pflamm-altenburg07a} or over $3-4$ crossing times \citep{tan06a, huff06, krumholz07e, jeffries07a, nakamura07, matzner07}. Since the latter is more restrictive, we adopt it. This leaves the problem of estimating the crossing time. To do so, we use a convenient form that expresses it in terms of a cloud's mass $M$ and column density $\Sigma$ \citep{tan06a}
\begin{equation}
\tcr = \frac{0.95}{\sqrt{\alpha_{\rm vir} G}}\left(\frac{M}{\Sigma^3}\right)^{1/4},
\end{equation}
where $\alpha_{\rm vir}$ is the cloud's virial ratio \citep{bertoldi92}, essentially its ratio of kinetic to gravitational potential energy up to geometric factors. Observed star-forming clouds all have $\alpha_{\rm vir}\approx 1-2$ \citep{mckee07b}, so we take $\alpha_{\rm vir}=1.5$ as typical, but $\Sigma$ is somewhat more difficult to estimate. Partly-embedded clusters detected in nearby star-forming regions using 2MASS and {\it Spitzer}/IRAC have surface densities of a few $0.01$ g cm$^{-2}$ \citep{adams06a, allen07a}, but because these samples are selected in the near-infrared they are strongly biased against more deeply-embedded and thus presumably younger clusters. Indeed, these samples are all dominated by stellar rather than gas mass, and thus have likely undergone significant expansion. 
Observations of gas-dominated cluster-forming regions show column densities of $\sim 0.1-3$ g cm$^{-2}$ 
\citep{plume97, mueller02, shirley03, faundez04a, fontani05a}; the data are summarized in Figure 1 of \citet{fall10a}.
We therefore adopt $\Sigma=0.3$ g cm$^{-2}$ as a fiducial value. This is toward the low end of the observed range, giving a high (and thus conservative) estimate of the crossing time. Together with an assumed formation time $\tform=4\tcr$ gives
\begin{equation}
\tform \approx 3.0 \left(\frac{\epsilon}{0.2}\right)^{-1/4} \left(\frac{M_*}{10^4\,\msun}\right)^{1/4}\mbox{ Myr},
\end{equation}
where $\epsilon=M_*/M$ is the star formation efficiency, i.e.\ the fraction of the initial cloud mass that is converted into stars. Observations suggest this is typically $10-30\%$ \citep{lada03}.

Since $\tform$ is the time required to form all the stars, we conclude that for clusters $10^4$ $\msun$ or smaller, almost all of the stars will form before the first supernova occurs within the cluster, so the gas will be chemically homogeneous. This is probably somewhat conservative, since only the most massive stars go supernova in 3 Myr, and such a massive star is unlikely to be found in a cluster containing only $10^4$ $\msun$ of stars. Moreover, $\tsn$ is the time delay between formation of a massive star and the resulting supernova, and a massive star is not necessarily the first to form in a given cluster. If we assume that massive star formation occurs on average halfway through the cluster formation process, then we only require $\tform < 2\tsn$ to ensure there is no self-pollution, which corresponds to a cluster mass $M_*=1.6\times 10^5$ $\msun$. We therefore conclude\footnote{It is interesting to note what happens in the case of globular clusters. These form at a density of  $\Sigma\sim 3$ g cm$^{-2}$ which leads to a sixfold drop in the dynamical time and a uniformity mass limit of $\gtrsim 10^7\msun$. Apart from a few light elements, globular clusters are chemically uniform, with few exceptions (e.g. $\omega$Cen; Gratton et al 2004).} that clusters smaller than 
$10^4$ $\msun$ should essentially all be chemically homogeneous, while a significant fraction of clusters up to $\sim 10^5$ $\msun$ will also be homogeneous.
This is to be expected if the
surface density of their more massive cluster-forming regions are
higher than for the lower-mass clusters.

\subsection{Unique chemical signatures}

The analysis in \S 3.2 allows us to derive an order-of-magnitude estimate of the number of unique chemical signatures
that are needed to identify stars with their parent clusters. Within the Solar toroid, the total number of tags required comes from integrating equation~\ref{eq:birth}  over the mass limits, viz.
\begin{equation}
\label{tags}
{{d{\cal N}}\over{dt}} = \int_{\ln M_{*,min}}^{\ln M_{*,max}} {{d^2 {\cal N}}\over{dM\; dt}} \; d(\ln M_*) = 6 \times 10^5\;  {\rm Gyr^{-1}}
\end{equation}

Thus, assuming that we could select stars with an age similar to the Sun ($t_o \approx 4.57 \pm 0.5$ Gyr), we would need of
order $6\times 10^5$ unique chemical signatures. Bland-Hawthorn \& Freeman (2004) suggested going after 
dozens of independent elements that could be measured accurately enough to distinguish a ``weak'' line
strength from a ``strong'' one. We can express this as $N_c = N_L^m$ where $N_L$ is the number of distinct line
strengths and $m$ is the number of independent elements or element groups. After detailed consideration
of what is possible in the context of a fibre-fed high-resolution spectrograph, the HERMES project has
settled on 8 independent groups with 5 measurably distinct abundance levels, i.e. $N_c \sim 5^8 \approx 4\times 10^5$ unique
chemical signatures. This is the target specification for the million-star Galactic Archaeology survey at 
the Anglo-Australian Telescope (AAT) due to commence in 2012 (Freeman et al 2010). 
The independent groups (e.g. alpha elements) include one or more elements that show some
variation with respect to [Fe/H].

The volume of the potential chemical space 
(${\cal C}$-space; Bland-Hawthorn \& Freeman 2004) accessible to the HERMES survey is barely adequate
for our proposed study. But further constraints will come from stellar ages at least for some stars. 
An important population is the subgiants for which ages can be derived from differential spectroscopy. 
In particular, a differential precision in surface gravity of $\Delta \log g \lesssim 0.1$ can be used to determine ages to $\sim$ 1 Gyr from
isochone fitting in the surface gravity vs. effective temperature plane.
There are some hidden assumptions in arriving at equation~\ref{tags} but we defer a discussion of these issues to a companion paper. 
However the analysis does emphasize the potentially vast amount of information that awaits us
in systematic high-resolution surveys of millions of stars.

\bigskip

\section{The Solar family}

Since the Sun is the star whose chemical abundances we know best, it is of particular interest for the purpose of chemical tagging. We would therefore like to estimate how many chemically-identical siblings the Sun might have. As we have
shown, this largely depends on the mass of the progenitor cluster.

Attempts have been made to constrain the size of the Sun's parent cluster using both both dynamical and nucleosynthetic evidence. Dynamically, \citet{adams01a} argue that the nearly circular orbits of the outer planets suggest that the Sun was born in a cluster with no more than $\msuncl \ltsim 10^4$ $\msun$. In a larger cluster the Sun would likely have experienced a close encounter that would have left Neptune with an elliptical orbit, and possibly stripped it entirely. However, as summarized by \citet{megeath08a}, this estimate is probably a bit low due to effects that \citeauthor{adams01a} did not include. Neptune probably migrated to its present position from closer to the Sun over a time scale of $\sim 10$ Myr, so during this period, when the encounter rate with other stars would be highest, the Solar System presented a smaller target than \citeauthor{adams01a} assumed. Moreover, even if the orbit of Neptune had been perturbed, its eccentricity might have damped due to interactions with a planetesimal disk. These effects imply that the dynamical constraint is weaker than Adams \& Laughlin's estimate $\msuncl \ltsim 10^4$ $\msun$, although by how much remains quite uncertain. We should add that on top of these planetary dynamics effects, there is considerable uncertainty in the \citeauthor{adams01a} calculation arising from the assumptions they are forced to make about the primordial binary fraction, the initial stellar density of the Sun's parent cluster, and how gravitationally unbound it was, all of which are quite uncertain.

The nucleosynthetic constraint comes from the presence of the daughter products of short-lived radionuclides in Solar System meteorites \citep{wadhwa07a}. These isotopes have half-lives in the range $0.1-3.7$ Myr (for $^{41}$Ca and $^{53}$Mn, respectively), so their equilibrium abundances in the interstellar medium are very small, well below the levels inferred from meteorites. Thus a local and non-equilibrium origin is required to explain their presence. Some of the isotopes, such as $^{10}$Be, were likely produced {\it in situ} in the early Solar System by x-ray irradiation \citep{shu97a, lee98a}, but others must be the result of stellar nucleosynthesis and injection of material into the proto-Solar System, either because that element is not produced efficiently by irradiation ($^{60}$Fe; \citealt{lee98a}), because the x-ray flux of young stars is insufficient to produce enough of it {\it in situ} ($^{7}$Be, $^{26}$Al, $^{26}$Cl, and $^{53}$Mn; \citealt{duprat08a}), or because timing constraints show that the element in question must have been injected in a single burst rather than produced over an extended period ($^{26}$Al; \citealt{thrane06a}; \citealt{connelly08a}). These elements must have an extrasolar origin, which is usually taken to be nucleosynthesis in a nearby massive star (although see \citealt{gounelle08a}, who argue for accretion of $^{60}$Fe and possibly $^{26}$Al from the ISM). 

The need for production in a nearby massive star sets constraints on the size of the Sun's parent cluster and its chemical homogeneity, albeit quite uncertain ones. There are three main scenarios for how supernova-produced elements might be incorporated into the gas out of which meteorites containing short-lived radionuclides eventually formed. First, a supernova explosion might have triggered collapse of a pre-stellar gas core a few pc away, leading to formation of the Sun in a gas cloud that has just been enriched by supernova ejecta \citep[e.g.][]{boss08a}. If this scenario is correct, then we have little hope of finding the Sun's siblings, because the supernova that injected the nuclides would also lead to significant chemical inhomogeneity in the Sun's parent cluster. Explaining the observed abundances of short-lived radio-nuclides requires that the pre-Solar core capture $\sim 10^{-4}$ of the total supernova ejecta \citep{meyer00a} and, as noted above in \S~\ref{clusterchemistry}, capturing even one part in $10^5$ of the supernova ejecta would be enough to change the iron abundance by a measurable amount. Thus stars that formed prior to the supernova would be chemically different than those that formed subsequently, and pre-stellar cores at different distances from the supernova blast wave would be enriched by different amounts. Thus $\msuncl$ would be large (probably $10^4-10^5$ $\msun$ or more in order to ensure a long enough cluster formation time to allow a supernova), but most of the stars would be chemically different from the Sun.

A second scenario is that the short-lived radionuclides were delivered via the winds of a Wolf-Rayet (WR) star rather than by a supernova \citep{arnould06a, gaidos09a}. The WR star would not have been part of the Sun's cluster, but would instead have been part of an earlier generation of stars born in the same giant molecular cloud (GMC), and would have contaminated that GMC. This scenario still requires supernovae to explain the presence of $^{60}$Fe, since this isotope is not present in significant quantities in Wolf-Rayet winds, but its relatively long half-life of 1.5 Myr, compared to 0.73 Myr for $^{26}$Al and 0.1 Myr for $^{41}$Ca, also allows the stars responsible for producing the contamination to have been part of an earlier generation rather than part of the Sun's parent cluster. Under this scenario we therefore have no constraint on the size of the cluster in which the Sun was born, since none of the contaminating stars were part of the Sun's cluster. Since the contamination in this scenario happens prior to formation of the Sun's cluster, there should be time for mixing to homogenize the contaminants and the uniformity of the cluster should not be compromised.

A third scenario incorporates the injection of short-lived radionuclides into the Sun's protoplanetary disk after star formation in the cluster is complete and the Sun's parent core has been fully accreted, but before both the cluster and the Sun's protoplanetary disk have dispersed (Ouellette et al. 2007).
Since in this case the supernova ejecta mix with a $\sim 0.01$ $\msun$ protoplanetary nebula rather than the entire $1$ $\msun$ pre-stellar core, the proto-Solar System need capture only $\sim 10^{-6}$ of the supernova ejecta to explain the observed abundance, and the chemical change induced by capturing such a small fraction of a supernova's ejecta is generally below the observationally-determined level of chemical homogeneity in star clusters. Moreover, in this scenario the Sun should be enriched much less than the proto-planetary nebula by the supernova, since its small cross section ensures that it captures very little supernova material directly, and only a small fraction of the enriched gas in the protoplanetary disk will accrete onto the Sun rather than being dispersed. Thus, this scenario is still compatible with the Sun being part of a chemically-homogeneous cluster.

We can estimate how large the Sun's parent cluster is likely to have been under the disk enrichment scenario based on the likelihood of a supernova occurring close enough to provide the observed level of enrichment and early enough to precede the disk's dispersal. Small clusters are unlikely sites for the formation of the Sun because they do not produce stars massive enough to go supernova before the protocluster disperses, while clusters that are too large are unlikely because they have short relaxation times that cause them to spread apart quickly. \citet{williams07a} and \citet{gounelle08a}, updating an earlier calculation by \citet{adams01a}, find that the probability of a supernova explosion enriching the Sun's protoplanetary disk to the degree required by observations is most likely in a cluster of $\sim 10^4$ stars, with a broad peak around that. (The absolute probability depends on the assumptions one makes about the initial cluster surface density, the duration of star formation, and the survival fraction of protoplanetary disks that come close to ionizing stars for at least parts of their orbits, but these do not change the most likely cluster size.) Given the uncertainties, we can state that under this scenario we expect the Sun to have been born in a cluster of $10^3-10^5$ $\msun$ in mass.

\bigskip

\section{In search of the Solar family}

In this section, we discuss basic parameters of the Solar family (e.g. its orbit properties) and of the Galaxy
disk to allow us to simulate the projected distribution of this particular group. From these simulations, we 
are led to possible survey strategies for finding group members.

\subsection{Solar parameters}
\label{Solarparam}

We know more about the Sun than any other star which makes it the best template for setting up a chemical tagging experiment. The Solar atlas of spectral lines includes 25,000 lines covering 66 elements (Moore, Minnaert \& Houtgast 1966; Wallace, Hinkle \& Livingston 1998). We know a great deal about the Sun, including its absolute abundances in all elements, its age, mass, luminosity, and its surface temperature. We know the depth of the convection zone, its central temperature, internal structure, and the strength of its magnetic field. We have some understanding of its surface activity,  surface structure and complex rotation. In principle, with access to a vast database of echelle spectroscopy, it should be possible to identify the Solar family more easily than any other disrupted star group, at least for Solar type stars.  This task will be greatly aided in an era of ESA {\it Gaia} when candidacy for the Solar family can be restricted by complete phase space information (see \S 7).

We assume the Sun was born in a dense cloud which gave rise to 10$^4$ stars to within some factor. A star cluster of this mass may have diffused within the last few billion years, but initially we shall assume that the star cloud has dispersed uniformly within a torus along the Solar circle.  Since proof of membership is obtained from chemical tagging (via heavy element abundances measured from high-resolution spectroscopy), we examine how many stars exceed apparent magnitude limits appropriate to modern telescopes, i.e. apparent magnitude limits in the range B = 14 to B = 20. 
These are the likely flux limits for a wide-field, multi-object spectrograph on any existing or foreseeable telescope.

\begin{figure}
\plotone{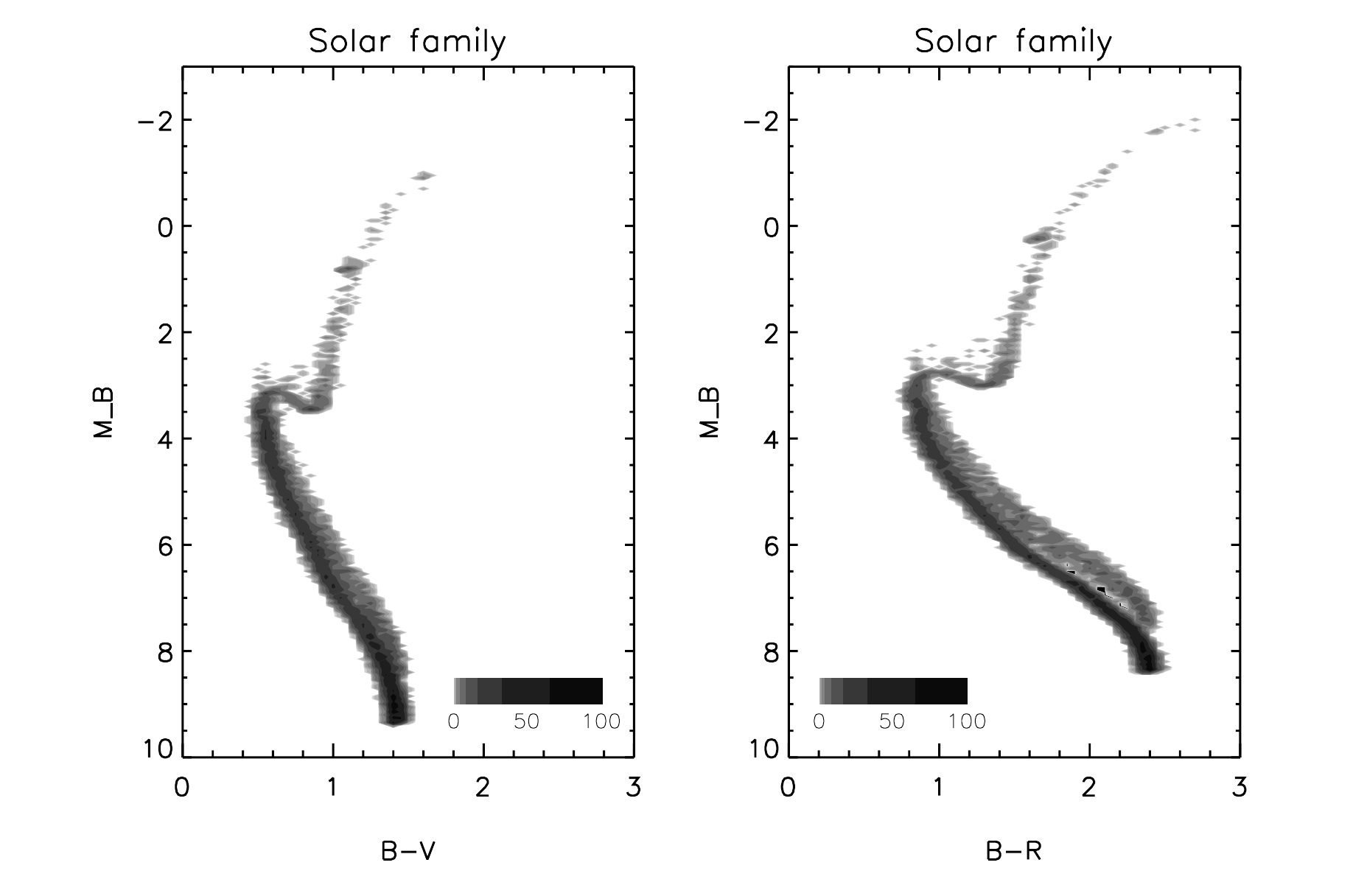}
\caption{ \label{cmd}
Predicted B-V and B-R colour-magnitude diagrams for the Solar family; the parallel main sequence arises from unresolved binary stars. The total population of simulated stars is $10^4$; the greyscale indicates the relative number of stars in bins of 0.05 along both axes.
}
\end{figure}

We compute the present-day colour-magnitude diagram (CMD) for the putative Solar family using the program 
StarFISH\footnote{See marvin.as.arizona.edu/$\sim$jharris/SFH}. We specify BASTI isochrones with the revised Solar metallicity $Z_o=0.0198$ (corresponding to [Fe/H]=0.06) for the known age of the Sun, 4.57 Gyr. A Salpeter IMF (logarithmic slope $-1.35$) is assumed with a 25\% binary star fraction.  Our total population is 10$^4$ stars with a minimum stellar mass of $0.1\msun$ and maximum mass of $100\msun$. StarFISH returns the data in pixelated form, for which the colour range sampling is (B-R and B-V): 0 to 3.0 ( in steps of 0.05); and magnitude range (R and V): 10 to $-3.0$ (in steps of 0.05).  The resulting CMDs are 
shown in Fig.\ \ref{cmd}.

We now consider the Sun's position in a cylindrical frame. 
The radius of the Solar circle is close to 8.0 kpc (Reid 1999). 
Gies \& Helsel (2005) have determined a useful orbit for the Sun within Dehnen \& Binney's (1998) realization of
the Galactic potential. The radial excursion keeps the Sun confined to the radial range ($R:7.8-8.6$) kpc essentially
consistent with an epicyclic approximation for a nearly circular orbit in an axisymmetric gravitational field (Lindblad 1959). If $\sigma_R$ is the velocity dispersion (in km s$^{-1}$) in the radial direction, and $\eta$ is the rms radial amplitude of the epicyclic motion in kpc, then $\sigma_R = 36\eta$. For the old stellar disk, $\sigma_R$ is about 40\kms\ (Nordstr\"{o}m et al 2004).

Thus the Sun is presently near pericentre and its guiding centre is at 200 pc beyond the Solar circle. Gies \& Helsel determine an excursion period close to the orbit period and a frequency that is essentially identical to the value deduced from Oort's constants ($36.7\pm2.4$ km s$^{-1}$ kpc$^{-1}$). While the Sun was long thought to be substantially offset from the disk plane, the {\it Hipparcos} astrometric satellite has convincingly demonstrated that its present position is $8\pm 4$ pc (Holmberg, Flynn \& Lindegren 1997). Gies \& Helsel determine a vertical oscillation amplitude of 140 pc roughly a factor of three less than what is expected for the old stellar disk (\S 5.3). Thus for the Solar toroid, we adopt a rectangular cross-section with radius 8.2 kpc, a radial thickness 800 pc and a vertical thickness of 140 pc (scale height 70 pc). In \S 6, we refer to this as Model 1.

\subsection{Galactic extinction}

In any projection of a dispersed stellar group, we must include the effects of interstellar extinction along the line of sight. We investigate this problem in two ways
which gave similar results. First, we generate a fractal medium on the scale of the Galactic disk. This is assumed to be 
a flat disk with a vertical thickness of 500 pc and with a total radial extent of 20 kpc. After Fischera \& Dopita (2005), we 
adopt a lognormal dust-gas distribution consistent with large-scale turbulence in a compressible medium. The
probability density distribution $P$ of the local density $\rho$ in the medium is given by
\begin{equation}
P(\rho) = {1\over\rho} P(\ln \rho) = {1\over{\sqrt{2\pi}\sigma_\rho \rho}} e^{-x^2/(2\sigma^2)}
\end{equation}
where $x = \ln \rho - \ln\rho_o$, $\sigma_{\rho}^2$ is the variance and $\rho_o$ is the mean of the lognormal density distribution.
If dust is a passive contaminant in this medium, then its local density distribution is the same as the gas density (Warhaft 2000). 
There are several methods for implementing such a medium: one of the most approachable is given by Elmegreen (1997)
but a more rigorous treatment uses the Fourier transform domain (Sutherland \& Bicknell 2007).
Fischera \& Dopita (2005) show how to normalize the synthetic medium to account for local extinction by the ISM at optical
wavelengths.

Our second method is to utilize the {\it IRAS} DIRBE extinction maps derived by 
Schlegel, Finkbeiner \& Davis (1998)\footnote{These authors supply a code for accessing the DIRBE maps at
http://www.astro.princeton.edu/$\sim$schlegel/dust
which we have
adapted for compilation under MAC OSX: a copy can be obtained from jbh@physics.usyd.edu.au.}.
At the present time, 
there is no 3D model of how the dust is distributed along each and every line of sight. We therefore adopt a statistical approach. The dust is assumed to be distributed uniformly along each line of sight. The total extinction $E$(B-V) is taken directly from DIRBE maps with a conversion to $A_{\rm B}$ taken from Table 2 in Schlegel et al (1998). The columnar extent of the dust distribution is assumed to be identical to the disk used in the above method. Our approach ensures that if we were to measure the extinction along each line of sight with an essentially infinite number of model stars, our synthetic extinction map would still match the complexity of the observed extinction map. 

\subsection{Radial and vertical diffusion}

When a star experiences deflections along its orbit, the phase space density around it changes with time.
It is well known that older stars have larger velocity dispersions than younger stars and, indeed, the study
of stellar diffusion has a long history (e.g. Spitzer \& Schwarzschild 1951; 1953). Binney \& Tremaine (2008;
hereafter BT) give the Schwarzschild equation which describes the velocity distribution of stars in the solar neighbourhood.
At a fixed age, the stars are described by a velocity ellipsoid with velocity dispersions $\sigma_R$, $\sigma_\phi$
and $\sigma_z$ with respect to the radial, azimuthal and vertical axis respectively. The axis ratios are roughly
constant as a function of stellar age but the volume of the ellipsoid grows with time, at least for a few Gyr
(Soubiran et al 2008; cf. Aumer \& Binney 2009).

A crude approach to understanding this behaviour is to assume {\it kinematic} diffusion 
using coefficients that are constant and isotropic (Wielen 1977; cf. Rosenbluth et al 1957). This model,
which is straightforward to implement within a numerical code, considers small velocity perturbations
$\Delta v_i$ and radial perturbations $\Delta r_i$ that have a mean of zero when averaged over all perturbations 
(denoted by the subscript $i$) over some period of time. But the
rms value need not vanish such that it increases with time $\Delta t$,
\begin{equation}
\langle \sum_i ( \Delta v_i)^2 \rangle = \delta_V \Delta t
\end{equation}
where $\delta_V$ is the velocity diffusion coefficient ($\delta_V = \delta_V(\delta_R,\delta_\phi,\delta_z)$).
This diffusion process also acts on the stellar positions such that
\begin{equation}
\langle \sum_i ( \Delta r_i)^2 \rangle = \langle \sum_i ( \Delta v_i (t-t_i))^2 \rangle
\end{equation}
where the perturbation at time $t_i$ led to the new position at time $t$.
For the observed vertical oscillation and in-plane epicyclic frequencies, $\omega_z$ and $\omega_{R\phi}$, the rms excursions are then obtained
in Galactocentric radius $R$ and vertical height $z$ from
\begin{eqnarray}
\langle (\Delta R)^2 \rangle &=& \int^t_o \delta_R\; {\rm sinc}^2 \omega_{R\phi} (t-t^{'}) \nonumber \\
& & +\; \delta_\phi (1-\cos \omega_{R\phi}(t-t^{'}))^2/(2B_{\odot})^2 dt^{'} \\
\langle (\Delta z)^2 \rangle &=& \int^t_o \delta_z\; {\rm sinc}^2 \omega_z (t-t^{'}) dt^{'}
\end{eqnarray}
for which\footnote{The sine cardinal function is defined as ${\rm sinc}(x) = x^{-1}\sin x$.} $B_{\odot}$ is Oort's constant.
Wielen (1977) considered three empirical relations for $\delta_V$: (i) a constant coefficient; (ii) a velocity 
dependent coefficient; (ii) a velocity-time dependent coefficient. We solve for each of these and present the results in Fig. 2.
The vertical dotted line indicates the age of the Sun and, as shown here, the characteristic scale height for a middle
aged star is 200-250 pc. The rms radial spread is $\sim$2 kpc, i.e. about a factor of three larger than realized by
Gies \& Helsel (2005).

But the velocity of most stars changes significantly along an orbit which argues
for a more sophisticated treatment. Indeed, Binney \& Lacey (1988) considered how stars diffuse in integral space
and showed that the diffusion coefficients vary from point to point. They considered Spitzer \& Schwarzschild's
proposal of random scatterings by giant clouds and found that, with the assumption of energy equipartition, $\sigma_z$ should
lie between $\sigma_\phi$ and $\sigma_R$. But this is not what is observed. The ratio of the last two quantities is
fixed by the epicyclic and azimuthal frequencies in the solar neighbourhood (BT; equation (8.117)), 
but $\sigma_z$ is observed to be smaller than both $\sigma_R$ and $\sigma_\phi$ (Nordstr\"{o}m et al 2004).
Ida, Kokubo \& Makino (1993) and Shiidsuka \& Ida (1999) challenged Binney \& Lacey's 
equipartition argument in a revised analysis and find that $\sigma_z$ is the smallest component of the dispersion, as observed.
This behaviour has been verified by Sellwood (2008) who demonstrates that the anisotropic distribution of the
(more distant) massive clouds determines the shape of the velocity ellipsoid.\footnote{Presumably this is strong evidence
against the recurrent claim of dark matter halos consisting of massive black holes.}

Some early papers (e.g. Carlberg 1987; Jenkins \& Binney 1990) appeal to transient spiral arms to assist with the 
vertical heating. But because the vertical oscillation frequency is higher than the radial epicyclic frequency, spiral 
scattering couples to vertical scattering only very weakly. Thus, the processes that drive vertical scattering are 
distinct from churning. In a recent twist, Sch\"{o}nrich \& Binney (2009) investigate the idea that at least
some of the vertical dispersion is assisted by the churning process; here, stars that have migrated from the inner disk 
retain their higher vertical dispersions when migrating outwards. But whether non-circular orbits can be efficiently
transported in this way remains unclear.

Aumer and Binney (2009) define an empirical velocity dispersion $\sigma^2$ for a known stellar distribution $N$ of a 
given age $t_o$ as
\begin{equation}
\label{aumer}
\sigma^2 = {{\int^{t_o}_0 \sigma^2(t) (dN/dt)\ dt}\over{\int^{t_o}_0 (dN/dt)\ dt}}
\end{equation}
where the vertical component is given by
\begin{equation}
\sigma_z(t) = 20 \left({{t+t_1}\over{t_D+t_1}}\right)^{0.38} .
\end{equation}
We use updated values for the coefficients (Binney 2010) where the age of the disk is $t_D =$ 10 Gyr
and for which $t_1=0.1$ Gyr. For our application, $t=t_o$ is the age (4.57 Gyr) of the Solar Family.

The vertical dispersion can be related to a physical scale, i.e. a characteristic scale
height or disk thickness, if the vertical component of the potential $\Phi_z$ is known since
${{1}\over{2}}\sigma_z^2 \approx \vert \Phi_z \vert$ (Kuijken \& Gilmore 1991). A more
rigorous treatment includes information from the vertical distribution function $f_z$, but the existing
data do not provide a consistent story (see Binney 2010). Suffice it to say, the mean scale height for
solar-type stars is close to $z_o \approx 250$ pc (Wielen 1977) which is confirmed by the more 
sophisticated analysis.

\subsection{Radial transport}

In the Wielen picture, star clusters dissolve and gradually diffuse into the background galaxy. Stellar populations
as an ensemble experience heating. The Sellwood-Binney mechanism is quite different in character
where perturbations allow the star to migrate in integral space without heating the disk.
During interaction with a single {\it steady} spiral 
event of pattern speed $\Omega_P$, a star's energy ($E$) and angular
momentum ($J$) change while it conserves its Jacobi integral. In the $(E, J )$ plane, stars 
move along lines of constant $I_J  = E - \Omega_P J$. For transient spiral perturbations, 
the star undertakes a random walk in 
the $(E, J )$ plane, deflected by a series of presumably uncorrelated spiral arm events which occur on time intervals
of 500 Myr or so. (The impulse acting on individual stars lasts for a fraction of this interval.)
So the Solar family would have experienced about $\sim$10 of these events
in the vicinity of co-rotation if only a fraction of the stars are moved in each event.

Substantial variations in the angular momentum of a star are possible over its lifetime. 
A single spiral wave near co-rotation can perturb the angular momentum of a star over a broad
distribution in $J_z$ with a tail up to 50\%.
The star is simply moved from one circular orbit to another, inwards or outwards, by up to 
2 kpc or so.  Conceivably, the same
holds true for a star cluster except that most are dissolved in much less than an orbit period. Given that there 
exists a long tail of very old star clusters, it is interesting to note that some of these could have been picked
up and moved to safety away from the major disturbances.

Support for the Sellwood-Binney mechanism (churning) comes from recent N-body simulations that allow
for a steady accretion of gas from the outer halo (Ro\v{s}kar et al 2008). These relatively ``cold'' simulations reveal 
some of the secular processes that may be acting over the lifetime of the disk. Averaged over the entire disk, 
these authors find that the rms radial excursion at $t=10$ Gyr is $\langle(\Delta R)^2\rangle^{0.5} \approx 2.4$ kpc and 
$\langle\Delta R\rangle \approx 0$ kpc. 
For the outermost disk stars,
the simulations show that $\langle\Delta R\rangle \approx 3.7$ kpc, several times larger than their
epicyclic radial excursion and our derived value of $\eta$ (\S5.1).

In order to understand the possible influence of churning on the Solar family, we envisage 
that at least one event affected the Solar family, i.e. those
that had the right pattern speed to move the members from wherever they were born to where they are 
today.  These events need not have treated all family members the same way. In all likelihood, the Solar family 
was already dispersed azimuthally at the time of the first event.  Depending on their orbital phase at the time, 
relative to the spiral event, some may have gone in and others out, and others may not have been much affected. 
There may have been subsequent events with the right pattern speed to move some members of the Solar 
family from their new radii to somewhere else. 

In order to model the impact of churning on the Solar Family, we use the illustrative formula provided by
SB02. The probability that a star formed at radius $R$ migrates to $R_o$ at time $t$ is given by
\begin{equation}
\label{SB02}
P(R_o\vert R) = (2\pi \sigma_t^2)^{-0.5} \exp\left(-{{(R_o-R)^2}\over{2\sigma_t^2}}\right)
\end{equation}
for which
\begin{equation}
\sigma_t = R_o ( (0.16)^2 + (0.4)^2 {{t}\over{t_o}})^{0.5}
\end{equation}
Using this simple formalism, we determine that the Solar Family could be dispersed over an annulus $\sim 5.5$ kpc.

Churning is a symmetric process enforced by angular momentum conservation. But a family
of stars in the solar neighbourhood can all move inwards together, or outwards, or even asymmetrically depending on how
much diffusion a star cluster has experienced and the consequent phase lags with respect
to the disturbing force (e.g. SB02, Figs. 4 and 13). 
Asymmetries can take place near inner (ILR) or outer Lindblad (OLR) resonances
in the sense that ILRs scatter inwards and OLRs scatter outwards.
Interestingly, both phenomena appear to have taken place in the solar neighbourhood in recent times.
Dehnen (2000) has invoked an outer Lindblad resonance driven by the central bar to explain the Hercules Stream.
Sellwood (2010) finds good evidence from the Hyades Stream for an {\it inner} Lindblad resonance which
presumably arose from a recent spiral disturbance that extended far out into the disk. This is evidence for the
phenomenon of transient spiral waves that propagate throughout the Galaxy due to large-scale disturbances
to the disk. In the simulations of Ro\v{s}kar et al (2008), there is clear evidence of asymmetry of
outward vs. inward transport at a radius of 8 kpc. But this reflects the limited extent of
the star formation beyond this radius.

\begin{figure}
\plotone{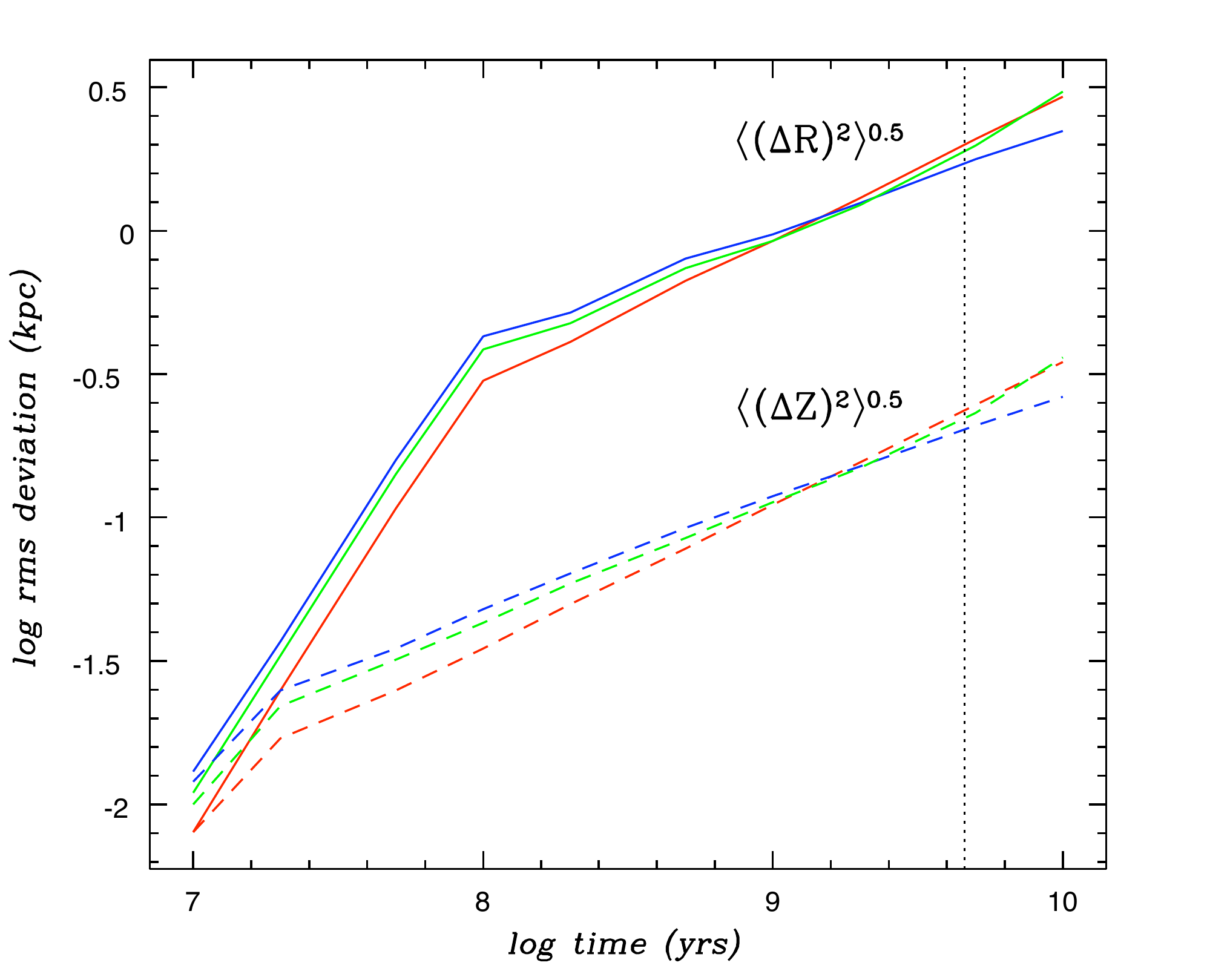}
\caption{ \label{diffuse}
Stellar diffusion in radius (solid curves) and vertical to the disk (dashed curves) as a function of cosmic time.
The different time-dependent relations (see text) were obtained from three different diffusion coefficients first derived by 
Wielen (1977).  The vertical dotted line indicates the age of the Sun.
\medskip
}
\end{figure}

\begin{figure}
\plotone{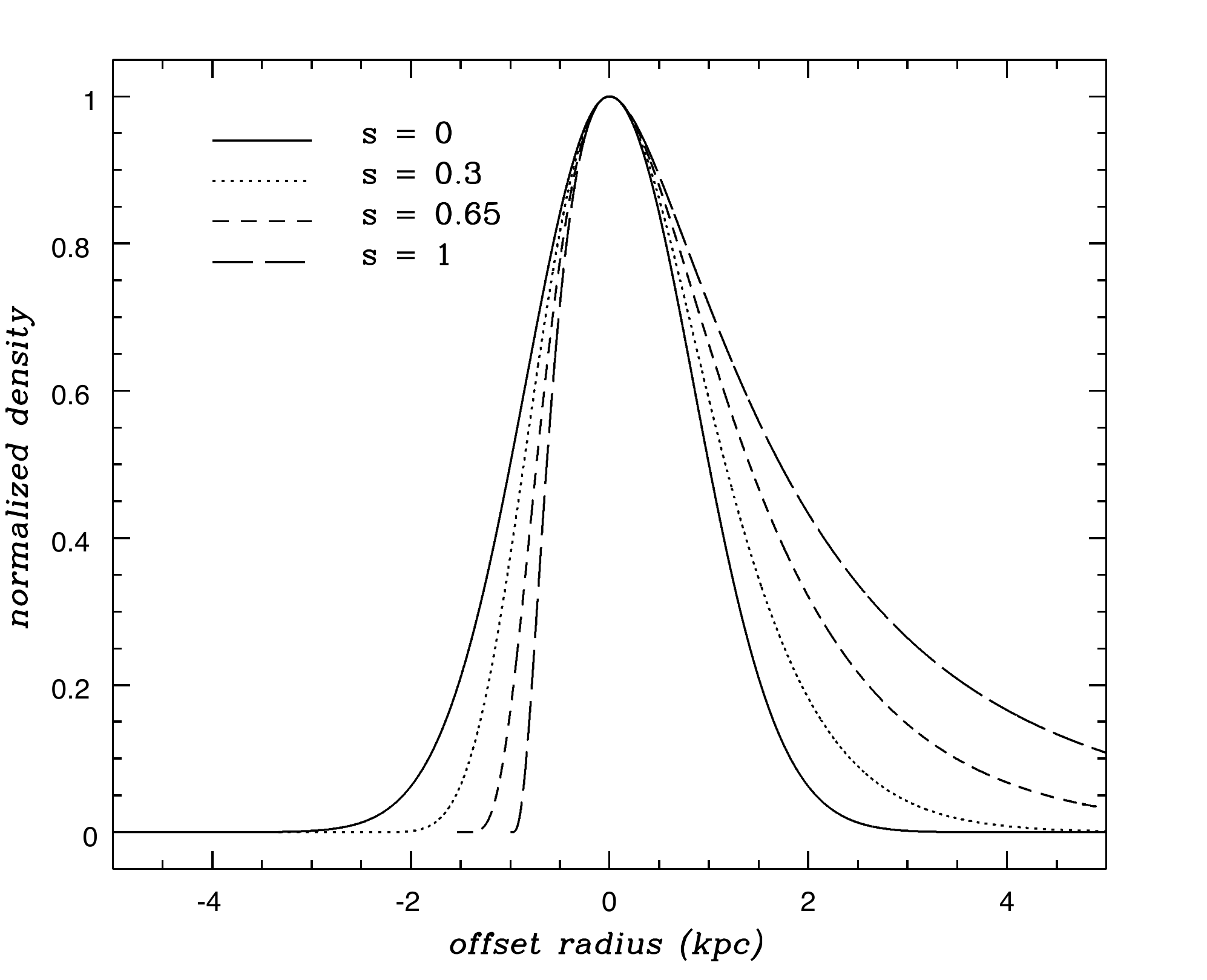}
\caption{ \label{skew}
The skewed gaussian function for four values of skewness: $s=0$ (solid), $s=0.3$ (dotted), $s=0.65$ (dashed),
$s=1.0$ (dot-dashed). These plots are normalized to the same peak density but can be normalized to the
total area under the curve (see text). This function is used to describe the cross-section of the Solar toroid for
different amounts of radial diffusion.
\bigskip
}
\end{figure}

To investigate the signature of a skewed distribution, we derive an empirical function
that has a skewed Gaussian cross-section in the radial direction.
We assume a coeval population that is distributed around the Solar toroid 
The Gaussian function has the form
\begin{equation}
G(R-R_o) = \rho_o \exp[-\ln 2 \cdot g(R-R_o)]
\end{equation}
where $g(x) = (2x/{\cal F}_g)^2$. The quantity ${\cal F}_g$ is the FWHM (i.e.  natural width) of the Gaussian toroid.
The cross-sectional area of the Gaussian toroid, $A_g = ({\cal F}_g/2)(\pi/\ln 2)^{0.5}$, allows the total volume of 
the toroid to be easily derived.

This is transformed to a skewed function by replacing $g(x)$ with $h(x)$ given by
\begin{equation}
h(x) = (\ln [1+2sx/{\cal F}_h]/s)^2
\end{equation}
where $s$ is a measure of skewness and $x > -{\cal F}_h/(2s)$. This has the nice property that 
$h(x)\rightarrow g(x)$ in the limit that $s\rightarrow 0$. We can therefore write
\begin{equation}
H(R-R_o) = \rho_o \exp[-\ln 2 \cdot h(R-R_o)]
\end{equation}
In Model 3 below (\S 6), the variable $s$ is time dependent. The quantity ${\cal F}_h$ is the FWHM (i.e. width) of the 
skewed Gaussian toroid which is a function of $s$. 
The cross-sectional area of the skewed toroid $A_h = ({\cal F}_h/2)(\pi/\ln 2)^{0.5}\exp[s^2/(4\ln 2)]$
allows the total volume of the toroid to be approximated or for the total area to be renormalized to the gaussian
cross-section with the same FWHM. In Fig.~\ref{skew}, we plot the function for $s=0, 0.3, 0.65, 1$.
Radial profiles with $s\sim 0.5$ resemble the slightly skewed histograms presented in Fig. 13 of SB02.
We emphasize that while certain annuli within the disk may exhibit this behaviour, churning conserves
angular momentum when summed over all sections of the disk.

\bigskip

\section{Results}

\subsection{Distribution in cylindrical coordinates}

\begin{figure}
\plotone{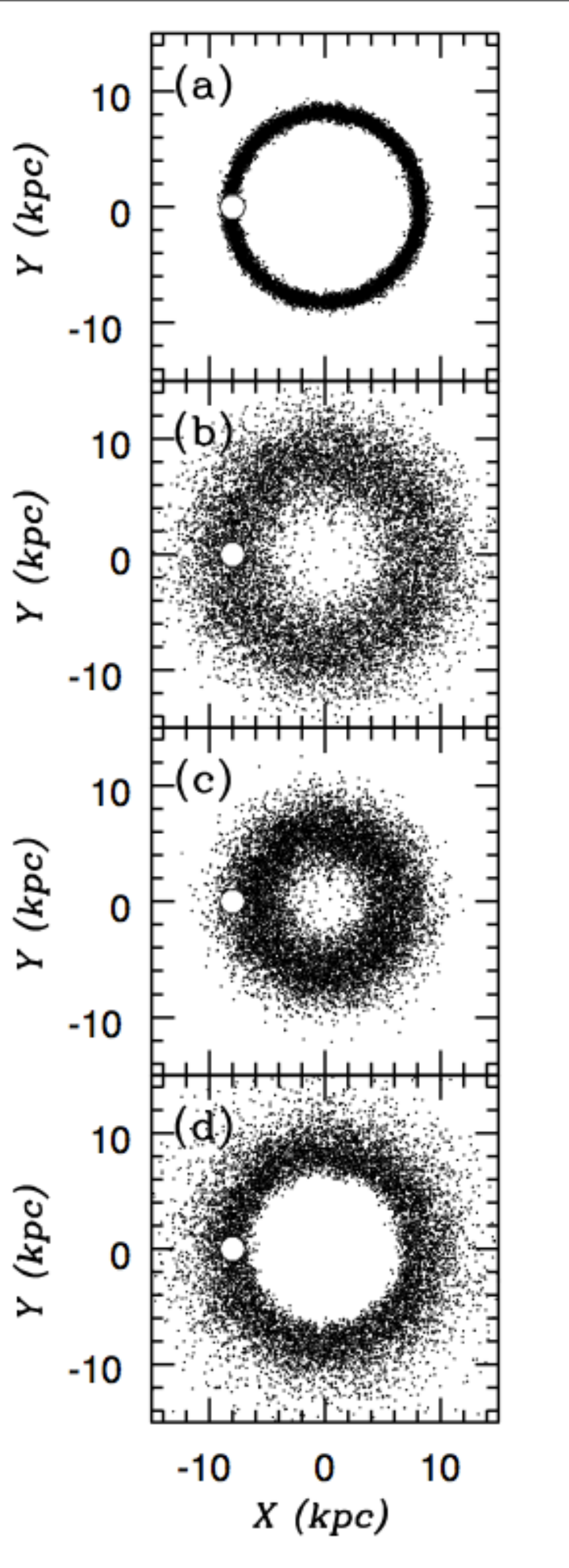}
\caption{ \label{annulus}
(a) Model 1: parameters from Gies \& Helsel (2005);
(b) Model 2: parameters from Sellwood \& Binney (2002);
(c) Model 3: parameters from Wielen et al (1996);
(d) Model 4: parameters to illustrate asymmetries observed in numerical simulations.
The white circle shows the Sun's present position.
}
\end{figure}

In Fig.~\ref{annulus}, we show four possible distributions for the Solar family today. The 
model parameters are inspired by different published studies:
\begin{enumerate}
\item toroid parameters derived from orbit modelling by Gies \& Helsel (2005) -- the radial and
vertical thicknesses (FWHM) are 800 pc and 140 pc respectively (see \S 5.1);
\item toroid parameters derived from equations 10-13 -- the radial and vertical thicknesses
(FWHM) are 5.5 kpc and 500 pc respectively (see \S 5.3 and \S 5.4);
\item toroid parameters derived from Wielen et al (1996) and equations 10-13  
where the stars are born at a radius
of $R=6$ kpc -- the radial and vertical thicknesses (FWHM) are 4.1 kpc and 500 pc respectively (see \S 5.3 and \S 5.4);
\item toroid parameters to illustrate a skewed distribution -- the vertical thickness (FWHM)
is 500 pc and the radial spread is asymmetric ($s=0.5$ at the present epoch; see \S 5.3 and \S 5.4).
\end{enumerate}
Note that in polar coordinates, the stellar density must be weighted by a factor of $R/R_o$ to
ensure a radial density that conforms to the distributions in Fig.~\ref{skew}.

\subsection{Distribution in Galactic coordinates}

In Figs. 5-8, we illustrate the distribution of Solar siblings in Galactic coordinates for the four
models in Fig. 4. Only stars above a magnitude limit $B=18$ are shown. The blue
box in each figure aids a comparison between the distributions. This region falls inside of the
LSR apex; we refrain from illustrating the mirror field that is symmetric about 
$(\ell,b)=(0,0)$. The distribution in longitude $\ell$ is quite different for each of the models.

Note that the blue box in all figures is centred at $\ell = 75^\circ$ rather than in the direction of
the Sun's motion. This is because we are
witnessing the integrated effect of the path length as the Solar torus bends inwards.
This is particularly apparent for Models 2 and 4. The high projected density of stars in Model 3 reflects the
higher surface density when projected onto the disk compared to Models 2 and 4.

The stars in Model 1 are characteristically nearer to the Sun; interestingly, useful numbers of
these stars are predicted to fall within a $B=14$ survey limit. Unsurprisingly, the stars
in Model 1 are also concentrated closer to the disk than in the other models which may make
them harder to observe in practice due to crowding. The offset position of the Sun in relation to the thin
torus (see Fig. 4(a)) leads to the asymmetric distribution of siblings on the sky at both low and high
galactic latitude, particularly in directions that look outside of the Solar Circle ($\ell: 90^\circ-270^\circ$).

The cumulative counts are presented in Fig.~\ref{appB} (all sky) and tabulated in Table 1. 
The counts specific to the box region ($\ell:60^\circ-90^\circ$) and neighbouring fields are also given in Table 1. 
We refer to longitude zone ($\ell:90^\circ-120^\circ$) as the outer region and longitude zone ($\ell:30^\circ-60^\circ$) as the
inner region. The trends observed in Table 1 between the inner and outer regions largely reflect what
we see in Figs. 5-8. Models 2 and 3 have a substantial inner population with a small fraction of outer
stars in the case of Model 3. Models 1 and 4 have a substantial outer population which undergoes
a reversal at faint magnitudes. Prima facie, models 2 and 4 look similar but the behaviour exhibited by the 
magnitude counts in Table 1 is different.

Our models are simply to illustrate the prospect of directly measuring
the impact of secular evolution within the disk.
We stress here that a more sophisticated approach to dust extinction is needed to be sure
of the number counts, particularly at the bright magnitude limit. This is particularly true for
Model 1 where the projected stars lie close to the Sun. We refer the reader to a
brief commentaries on this issue by Robin (2009) and by Aumer \& Binney (2009).

\subsection{Survey strategy}

As we saw in the last section, the distribution in $z$ and the centroid position in longitude are 
powerful discriminants for determining the Sun's origins and the importance of in-plane diffusion 
and transport.  We now briefly consider possible survey strategies for four different apparent magnitude
limits. We restrict our discussion to the B band because this spectral region provides the richest information
for chemical tagging (Bland-Hawthorn \& Freeman 2004). For all of the strategies below, it should be
possible to achieve the required number of chemical tags derived in \S3.3.

In Fig.~\ref{appB}, we show four possible B magnitude limits (indicated by the vertical lines)
for high-resolution spectroscopy. The canonical values are as follows:
\begin{itemize}
\item B=14 is an appropriate limit on a 4m class wide-field spectrograph operating in 2\arcsec\
seeing. This is a magnitude brighter than the effective limit of the HERMES instrument (V$\sim$14)
that sees first light at the AAT in 2012.
\item B=16 is a suitable limit for a HERMES-style instrument carrying out a shallow survey
on an 8m class telescope in 1\arcsec\ seeing.
\item B=18 is an appropriate limit for a HERMES-style instrument carrying out a deep survey
on an 8m telescope in 1\arcsec\ seeing.
\item B=20 is a suitable limit for a HERMES-style instrument undertaking a million-star survey on an
extremely large telescope (ELT) in 1\arcsec\ seeing.
\end{itemize}

Table 1 indicates the number of Solar Family stars expected to be found in a million-star survey within the blue box region for different magnitude limits for Models 1-4. Notice that in each of the models, the maximum number of Solar siblings is detected for surveys with a limiting magnitude of B = 16-18. This peak is a result of the competition between two effects. For limiting magnitudes of B = 16 or less, only the brightest few Solar siblings are included in the sample, so increasing the limiting magnitude increases the number of Solar siblings found. However, the total number of stars in the survey field that are not Solar siblings also increases as one goes to fainter limiting magnitudes, and thus as the limiting magnitude B increases, the fraction of surveyed stars that are Solar siblings begins to decrease. Beyond B = 18, this decrease in the fraction of stars that are Solar siblings outweighs the increase in the absolute number of Solar siblings brighter than the magnitude limit, and the expected number of Solar siblings discovered in a survey of fixed size therefore decreases as the magnitude limit gets fainter. Thus the optimal magnitude limit of B = 16 to B = 18 is determined by a combination of the distribution of Solar family B magnitudes, illustrated in Fig.~\ref{appB}, and the distribution of B magnitudes for all stars in the Galaxy, which rises sharply above B = 18. 
The brighter limit is to be
preferred because of the higher signal and because of the reduced contamination from fainter stars
close to the Galactic plane.

A reasonable strategy is to survey the blue box region and control fields of similar size on the other
side of the LSR apex (see columns 2 and 4 in Table 1). The longitude test may be
relatively free from identification errors although a possible complication is the colour
selection of the input catalogue.  The more numerous faint stars with lower apparent
magnitudes will be more reddened than brighter stars on average since they will be
stars that are embedded deeper in the torus. However, stars that are far from the Solar
torus in longitude will be less reddened. A selection criterion will need to take this into
account. In an era of ESA {\it Gaia}, with the advent of proper motions and 6D phase space information for millions of stars, 
this test can be conducted with higher sensitivity in a physical coordinate frame.

We refrain from a detailed analysis of survey strategies at this time because there
are many issues to consider in an observing campaign. The model distributions themselves 
impose their own challenges. If the Gies \& Helsel solution for the Solar orbit is correct, the vertical 
concentration will keep the stars much closer to the Galactic plane. This has two problems: 
higher galactic extinction on average and stellar crowding.  In fact, for
bright stars this may not be a serious problem even in the plane. 
The APOGEE survey is targetting red clump stars and red giants 
down to H$\approx$13  between $b=-15^{\circ}$ and
$b=15^{\circ}$ for the entire longitude range observable from the northern
hemisphere (Allende Prieto et al 2008). This survey makes
use of the SDSS 2.5m telescope under natural seeing conditions. The chemical information 
appears to be limited at the present time although this is currently under investigation
(J. Lawler, personal communication). The H-band limit is equivalent to roughly B$\approx$16 and so 
somewhat deeper than the HERMES survey at the AAT3.9m telescope that will target spectral windows in the
V and R bands.

In the search for the Solar family, dense stellar fields can be an advantage with wide-field AO systems 
and multi-object spectrographs. 
In particular, one can comfortably envisage high-resolution IR spectrographs operating with multi-conjugate 
adaptive optics in the next decade, either on 8m class telescopes or on ELTs.
The ``echidna" positioning technology permits 4000 stars to be observed simultaneously
as was originally proposed WFMOS instrument (Colless 2005). Another related actuator technology provides the basis
for the LAMOST project (Zhang \& Qi 2008). These new developments can be extended to even larger
formats on wide-field focal planes. It is reasonable to suppose
that such a facility linked to an AO system will become available in the next decade because {\it inter alia} 
some of the most 
ancient stars are expected to have settled to the inner bulge (Bland-Hawthorn \& Peebles 2006; Tumlinson 2010).

\section{Discussion}

Within the domain of Solar physics, planetary science and geophysics, the formation and 
evolution of the Solar nebula is of paramount importance. This arena has been revitalised
by the discovery of exoplanets around nearby stars. In certain important respects, a search
for the Solar family is the ``missing link'' between top-down models of Galaxy disk formation,
and bottom-up models of Solar system formation. The search for the Sun's 
siblings is the first step to placing its progenitor cloud in
the context of the interstellar medium and the Galaxy as a whole.

We have seen that a reasonable number of the Sun's siblings can be identified in a targetted
million-star survey, at least in principle. Given that strong resonances have clearly
influenced the solar neighbourhood at different times (Dehnen 2000; Sellwood 2010), it 
would be surprising if the Solar family has escaped significant migration into the 
surrounding disk. Even more surprising would be the failure to identify a single family member.

Indeed, we would argue that chemical tagging, or a related technique, is
demanded by secular processes like the Sellwood-Binney mechanism. If this process
is as strong as implied by recent simulations, then most stars may not have formed {\it in situ},
which is an axiom of chemical evolution models. In such an event, a certain degree of
reconstruction is called for although much can be learnt from the higher moments of
the chemical space. The fact that ancient star clusters like Collinder 261 (De Silva et al 2007) 
have survived this process may indicate that these were picked up and transported
early in their history out of harm's way.

In \S3.3, we argued that full-blown chemical tagging requires $N_c \sim 6\times 10^5$ unique tags
per Gyr which may be within reach of the HERMES survey in its present incarnation.
But here the impending impact of the {\it Gaia} astrometric mission becomes clear. 
The Sun's orbit is one of the more
circularized of nearby dwarfs (Robles et al 2008). Relative to other
stars of similar age, its peculiar velocity is in the lowest 8\% of a Maxwellian distribution,
consistent with the orbital eccentricity derived by Robles et al (2008).
The Sellwood-Binney ``churning'' process does not impart any random energy (i.e. heating).
Thus, accurate phase space information from {\it Gaia} once available can allow us to 
preselect orbit families, for example, stars with circularized orbits. The bright end of this
distribution (say V$<$14)  then provides the primary input catalogue for a chemical tagging 
survey where {\it Gaia} is expected to yeld its most accurate phase space information.

The use of phase space information to aid the reconstruction of ancient star clusters is fraught
with problems at the present time. The success or failure of this approach relies on the assumption that the Sun's low peculiar velocity will be shared by all the other stars in the Solar family. Whether this is true or not depends on whether the Sun has a low peculiar velocity simply because random diffusion processes just happened to put it at the low velocity tail of the distribution, or because of some property of its parent cluster that will be shared by all stars that were born in it. This uncertainty is
partly reflected in our choice of three different models for the Solar family distribution based on very different assumptions.
Clearly, some knowledge of the distribution of the Solar family will go a long way to resolving these issues, and the
importance of secular processes in general.

In our attempt to recover the Solar family, we will identify many
thousands of distinct stellar groupings in chemical space.
In our companion paper, we derive the expected number of distinct stellar groupings and 
the inferred progenitor mass distribution (Sharma et al, in preparation). 
Of these long-dissolved star clusters, some will have much better representation, and many
will have too few members for a reliable identification. We can construct the colour-magnitude
diagram of candidate groupings to see if they are consistent with a coeval population.

Just how the reconstructed ancient cluster mass distribution will help to unravel the 
processes of Galaxy formation will
be investigated in future papers. In large part, it will depend on the detailed structure of
the ${\cal C}-$space, of which very little is known. There is widespread evidence of
anomalous chemical signatures in many discrete stellar systems (De Silva et al 2008). Just how these
signatures relate to the multi-dimensional topology of the chemical space will require 
much larger high-resolution spectral surveys to answer properly.

At the present time, it is unclear what constitutes the true building blocks of galaxies, 
whether these are early-stage dwarf galaxies (i.e. the progenitors of those that orbit the Galaxy)
or simply gas clouds confined by
infalling dark-matter halos. But the newly discovered ``ultra-faint'' dwarf galaxies (Simon \& Geha 2007)
may also be an important constituent of the formation process. These may carry unique chemical
signatures of the first stars. The faint end of the dwarf galaxy luminosity function can only be
studied within a few megaparsecs of the Galaxy. Many of these systems may have already
dissolved within the Galaxy to be revealed in future ``chemical tagging'' and astrometric
surveys.

It is clear that we are only now beginning to understand just how much information
is locked up in stellar atmospheres. In the ESA {\it Gaia} era, we will presumably
identify tens of thousands of stellar groupings in phase space or in integral space. 
Many of these may be spurious dynamical groupings created by spiral arm density
waves (De Simone, Wu \& Tremaine 2004; Quillen \& Minchev 2005). But many of these may be
disrupted ancient star clusters, some formed within the Galaxy, some external to it.
Some of these systems may even carry the hallmarks of the first generations of stars
in the early universe (Tumlinson 2010). Once again, chemical tagging will provide
key information on the likely origin of these low-mass structures.

If the chemical tagging technique is ultimately successful, it will provide
rigid constraints on the importance of the Sellwood-Binney mechanism and,
by implication, on the importance of secular processes and the lifetime of spiral arms. These are fundamental properties of
galaxies that must be understood if we are to have any hope of claiming a
detailed physical understanding of galaxy evolution.

\begin{figure}
\plotone{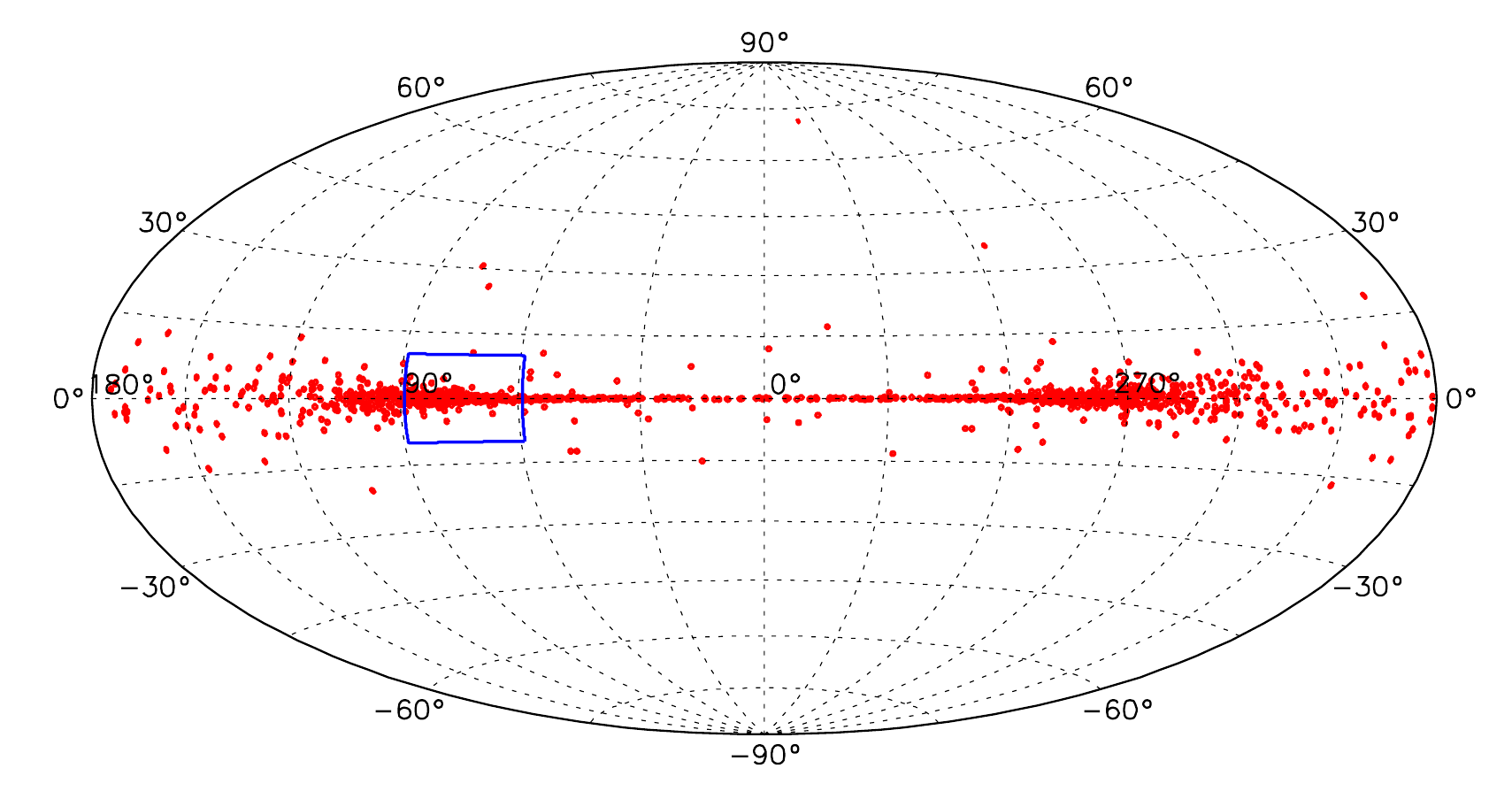}
\caption{ \label{sky1}
Model 1 distribution of Solar family stars brighter than B=18 (see Fig. 4(a)) in Galactic coordinates as seen from the Sun. 
Note that the stars are distributed symmetrically about the LSR apex and antapex in longitude,
quite unlike what is seen in the other models. Of the four models presented in Figs. 5-8, this model
is the most susceptible to how we treat the local dust distribution (cf. Robin 2009).
The blue box is the survey region discussed in the text.
}
\end{figure}

\begin{figure}
\plotone{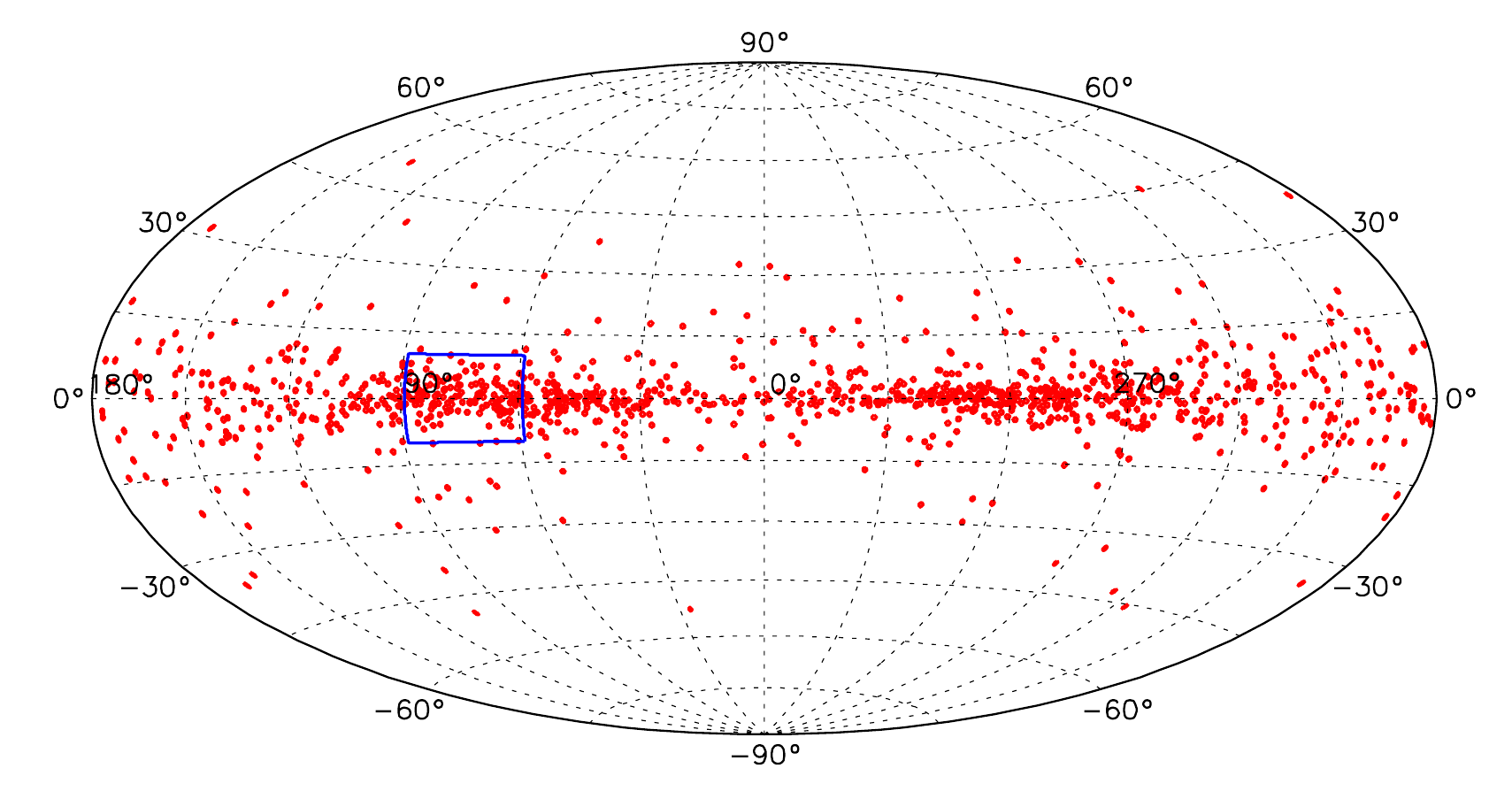}
\caption{\label{sky2}
Model 2 distribution of Solar family stars (see Fig. 4(b)) in Galactic coordinates brighter than B=18.
The blue box is the survey region discussed in the text.
}
\end{figure}

\begin{figure}
\plotone{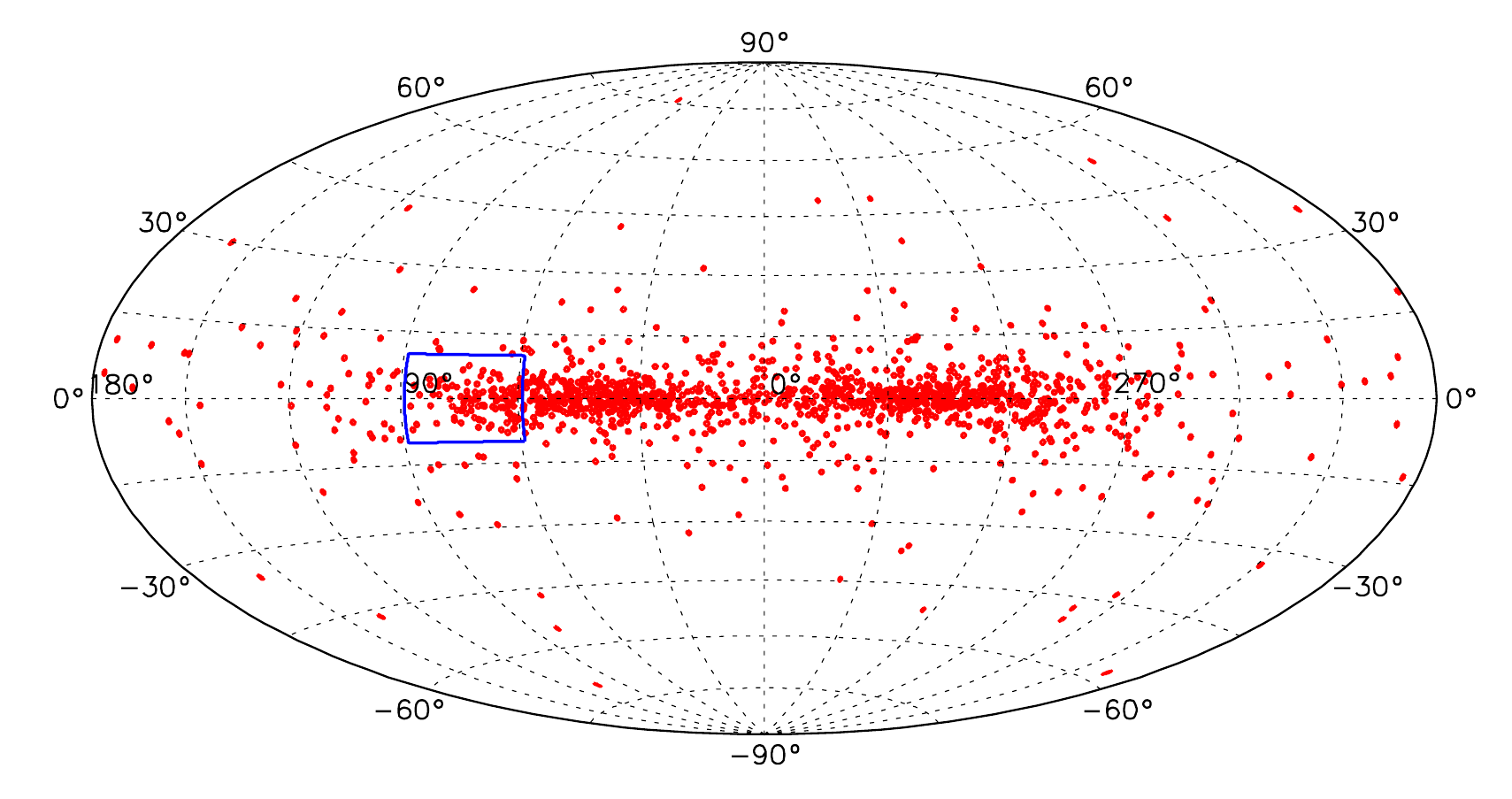}
\caption{\label{sky3}
Model 3 distribution of Solar family stars (see Fig. 4(c)) in Galactic coordinates brighter than B=18. Most of the stars fall ``inside'' of the 
galactic latitude (projected radius)
of the blue box, discussed in the text, with few stars outside.
}

\end{figure}
\begin{figure}
\plotone{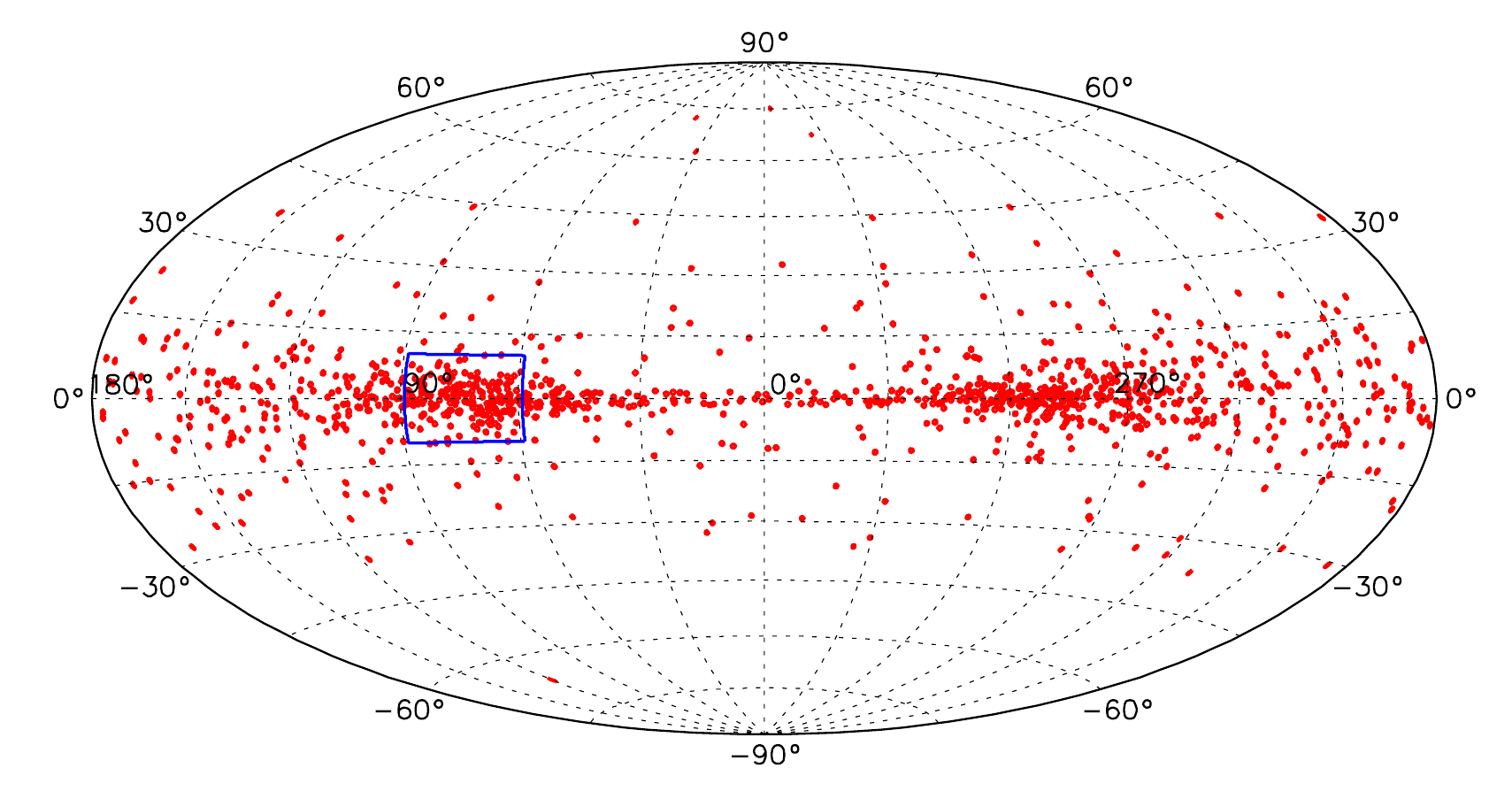}
\caption{\label{sky4}
Model 4 distribution of Solar family stars (see Fig. 4(d)) in Galactic coordinates brighter than B=18. Most of the stars appear highly localised in two clouds each covering about 300 sq. deg. as illustrated by the blue box at $\ell=60^\circ$ to $\ell=90^\circ$.}
\end{figure}

\begin{figure}
\plotone{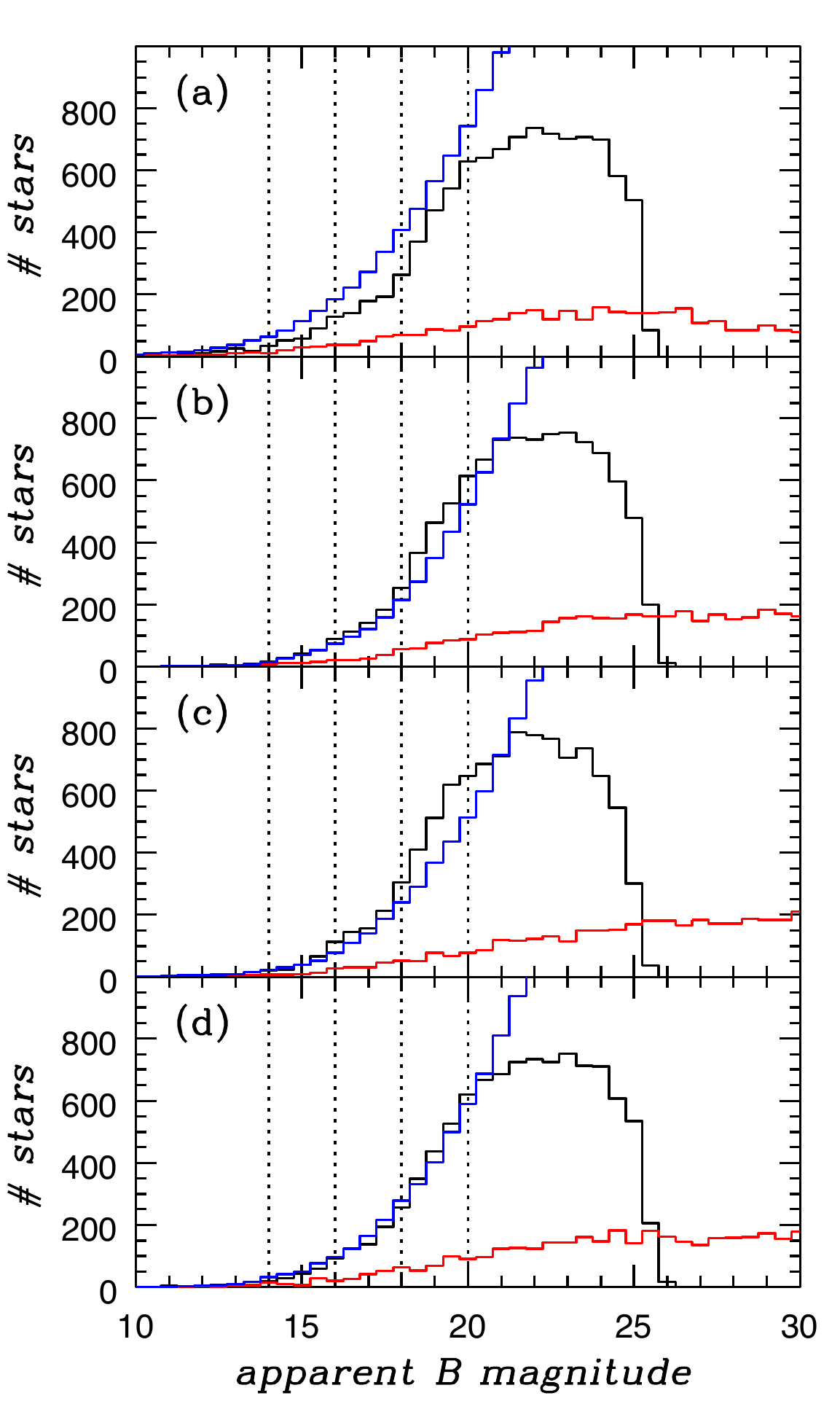}
\caption{ \label{appB}
Stellar number counts for (a) Model 1 (b) Model 2 (c) Model 3 (d) Model 4; see Fig. 4.
The black histogram shows the number of stars for each bin in apparent B magnitude  before dust correction, red shows the number of stars in each B magnitude bin after dust correction, and blue shows the cumulative distribution of stars (corrected for extinction) with apparent magnitude brighter than B mag. The vertical dashed lines indicate the magnitude limits for four
different surveys -- see text.
}
\end{figure}

\begin{table}
\caption{Cumulative B magnitude-limited star counts for the models illustrated in Fig. 4. Column 1 is the
apparent B magnitude limit. Column 2 gives the star counts inside of the blue
box centred at $(\ell,b)=(105^\circ,0)$ for magnitudes brighter than the B limit in Figs. 5-8.
Columns 3 and 4 are the same as column 2 but centred at $(\ell,b)=(75^\circ,0)$ and $(\ell,b)=(45^\circ,0)$.
Column 5 gives the star counts 
for the entire sky brighter than the B limit. Column 6 is the predicted
number of Solar siblings in a million star survey down to the B limit.}
\begin{center}
\begin{tabular}{cccccc}
\\
{\bf B mag} & {\bf 90$^\circ$:120$^\circ$} & {\bf 60$^\circ$:90$^\circ$} & {\bf 30$^\circ$:60$^\circ$} & {\bf total sky} & {\bf ${\bf 10^6}$ survey}\\
\\
{\it Model 1} & & & & \\
\\       
        14 &        18 &        10 &         6 &       113 &        14 \\
        16 &        53 &        56 &        18 &       401 &        36 \\
        18 &       119 &       271 &        55 &      1096 &        26 \\
        20 &       200 &       614 &       332 &      2919 &        14 \\
\\
{\it Model 2} & & & & \\
\\ 
        14 &         4 &         3 &         0 &        36 &         4 \\
        16 &        16 &        14 &        11 &       214 &        10 \\
        18 &        58 &       105 &        90 &       824 &        11 \\
        20 &       140 &       293 &       379 &      2613 &         7 \\
\\
{\it Model 3} & & & & \\
\\
        14 &         1 &         4 &         7 &        50 &         3 \\
        16 &         3 &        20 &        29 &       240 &         8 \\
        18 &        17 &        85 &       171 &       955 &         7 \\
        20 &        40 &       185 &       590 &      2975 &         4 \\
\\
{\it Model 4} & & & & \\
\\
        14 &         3 &         4 &         1 &        51 &         4 \\
        16 &        19 &        20 &         5 &       231 &        13 \\
        18 &        68 &       142 &        57 &       860 &        14 \\
        20 &       155 &       382 &       308 &      2588 &         9 \\
\end{tabular}
\end{center}
\label{tab}
\end{table}

\acknowledgments  JBH is supported by an Federation Fellowship from the Australian Research Council (ARC).
JBH and KCF are assisted in part by ARC grant DP0988751 that provides partial support for the HERMES project.
MRK is supported by the National Science Foundation through grant AST-0807739; by NASA through the Spitzer 
Space Telescope Theoretical Research Program, provided by a contract issued by the Jet Propulsion Laboratory; 
and by the Alfred P. Sloan Foundation through a Sloan Research Fellowship.
JBH and MRK would like to thank David Gross and the Kavli Institute at UC Santa Barbara 
for their hospitality during a critical period of this work. The Kavli research programmes are supported by
National Science Foundation grant PHY05-51164. We are also indebted to Matthias Steinmetz, Juna Kollmeier
and Andrew Benson for organizing the extended workshop. JBH thanks Merton College and the BIPAC Institute
at Oxford for their hospitality during the latter stages of this work. We are indebted to Jerry Sellwood for his detailed referee
report that helped to clarify aspects of this work, and we acknowledge helpful conversations with James
Binney, Sanjib Sharma and Ralph Sch\"{o}nrich.

\bibliographystyle{apj}

\end{document}